\documentclass[twocolumn]{aastex631}

\shorttitle{Dynamical Viability Assessment for Habitable Worlds Observatory
  Targets}
\shortauthors{Stephen R. Kane et al.}

\begin{document}

\title{Dynamical Viability Assessment for Habitable Worlds Observatory
  Targets}

\author[0000-0002-7084-0529]{Stephen R. Kane}
\affiliation{Department of Earth and Planetary Sciences, University of
  California, Riverside, CA 92521, USA}
\email{skane@ucr.edu}

\author[0000-0002-4860-7667]{Zhexing Li}
\affiliation{Department of Earth and Planetary Sciences, University of
  California, Riverside, CA 92521, USA}

\author[0000-0002-0569-1643]{Margaret C. Turnbull}
\affiliation{SETI Institute, Carl Sagan Center for the Study of Life
  in the Universe, Off-Site: 2613 Waunona Way, Madison, WI 53713, USA}

\author[0000-0001-8189-0233]{Courtney D. Dressing}
\affiliation{Department of Astronomy, 501 Campbell Hall \#3411,
  University of California, Berkeley, CA 94720, USA}

\author[0000-0001-5737-1687]{Caleb K. Harada}
\altaffiliation{NSF Graduate Research Fellow}
\affiliation{Department of Astronomy, 501 Campbell Hall \#3411,
  University of California, Berkeley, CA 94720, USA}


\begin{abstract}

Exoplanetary science is increasingly prioritizing efforts toward
direct imaging of planetary systems, with emphasis on those that may
enable the detection and characterization of potentially habitable
exoplanets. The recent 2020 Astronomy and Astrophysics decadal survey
recommended the development of a space-based direct imaging mission
that has subsequently been referred to as the Habitable Worlds
Observatory (HWO). A fundamental challenge in the preparatory work for
the HWO search for exo-Earths is the selection of suitable stellar
targets. Much of the prior efforts regarding the HWO targets has
occurred within the context of exoplanet surveys that have
characterized the stellar properties for the nearest stars. The
preliminary input catalog for HWO consists of 164 stars, of which 30
are known exoplanet hosts to 70 planets. Here, we provide a dynamical
analysis for these 30 systems, injecting a terrestrial planet mass
into the Habitable Zone (HZ) and determining the constraints on stable
orbit locations due to the influence of the known planets. For each
system, we calculate the percentage of the HZ that is dynamically
viable for the potential presence of a terrestrial planet, providing
an additional metric for inclusion of the stars within the HWO target
list. Our analysis shows that, for 11 of the systems, less than 50\%
of the HZ is dynamically viable, primarily due to the presence of
giant planets whose orbits pass near or through the HZ. These results
demonstrate the impact that known system architectures can have on
direct imaging target selection and overall system habitability.

\end{abstract}



\section{Introduction}
\label{intro}

Although thousands of exoplanets have now been discovered, a full
census and exploration of nearby planetary systems has yet to be
realized. One of the primary reasons for incompleteness regarding
nearby planetary architectures are the observational biases inherent
within the two detection techniques responsible for the vast majority
of current exoplanet discoveries: the radial velocity (RV) and transit
methods
\citep{cumming2008,kane2008b,ford2014,winn2015,kipping2016d,he2019}. Both
of these methods are heavily biased toward large, close-in planets,
with strong dependencies on survey design, data precision, observing
strategy, and the time baseline of observations
\citep{kane2007a,ford2008a,vonbraun2009,wittenmyer2013a,barclay2018}. However,
RV surveys that have been operating for multiple decades have been
able to probe beyond the snowline, providing insight into the
prevalence of giant planets, even at relatively large separations from
the host star \citep{wittenmyer2020b,fulton2021,rosenthal2021}. The
architectures of these systems are informative for their orbital
dynamics and evolution, that can also be used to constrain the
possible presence of additional planets within those systems
\citep{barnes2004b,kopparapu2010,kane2015b,kane2019e,kane2023c,kane2023d}.
Of particular interest is the potential for harboring planets within
the Habitable Zone (HZ) of the system
\citep{kasting1993a,kane2012a,kopparapu2013a,kane2014a,kopparapu2014,kane2016c,hill2018,kane2020b,hill2023}.
Such HZ planets are expected to be the target of extensive follow-up
observations, especially those nearby systems whose host star
brightness and angular separation of their planets may enable direct
imaging as a means to characterize the planetary atmospheres
\citep{brown2015a,barclay2017,kane2018c,kopparapu2018,stark2020,li2021a}.

Among the recommendations of the 2020 Astronomy and Astrophysics
decadal survey (hereafter Astro2020) was the prioritization of a
space-based mission with the capability to directly image terrestrial
planets within the
HZ \footnote{https://www.nationalacademies.org/our-work/decadal-survey-on-astronomy-and-astrophysics-2020-astro2020}. The
currently adopted name for this recommended mission is the Habitable
Worlds Observatory (HWO), and considerable efforts are being
undertaken to define the mission designs that would maximize the
science yield for terrestrial planet characterization
\citep{vaughan2023,stark2024}. Work is also being carried out to both
define and characterize the target stars that are best suited for HWO
observations. An initial list of 164 HWO target stars has been
provided to the broader community for assessment and discussion
\citep{mamajek2024}, the stellar properties of which have been further
refined by \citet{harada2024b} and examined in the context of yield
simulations by \citet{tuchow2024}. Of these 164 nearby stars, all of
which lie within 24~pcs, 30 are currently known to host exoplanets,
with a total inventory of 70 known exoplanets. These exoplanet
detections have primarily been enabled through long-term RV
observations, revealing the Keplerian orbits and architectures down to
the RV precision of the corresponding surveys. Though the
architectures for these planetary systems are likely incomplete, even
partial knowledge of the planets present enables a dynamical analysis
that can inform the likelihood that each system has a HZ that contains
dynamically viable orbital locations for terrestrial planets. Given
the need to determine a high quality list of HWO targets,
deprioritizing those systems that are dynamically unsuitable is a
crucial step in the target assessment process.

In this work, we present the results of a dynamical study for the
known exoplanet hosts within the current proposed HWO target list
provided by \citet{mamajek2024}, that specifically provides an
assessment for the dynamical stability within the HZ of those
systems. Section~\ref{sample} provides a description of the HWO
stellar sample, with a particular emphasis on the known exoplanetary
systems within that list. The description includes a tabulation of the
stellar and planetary properties for the known exoplanet systems, and
the methodology for the calculation of the system HZ boundaries. The
structure and results from the dynamical simulations are described in
Section~\ref{dynamics}, with a quantification of the available HZ that
is dynamically stable. We discuss the implications and limitations of
our work in Section~\ref{discussion}, and provide concluding remarks
and suggestions for future work in Section~\ref{conclusions}.


\section{The Habitable Worlds Observatory Sample}
\label{sample}

Here, we provide the details of the potential HWO target list that
harbor known exoplanets, and discuss the calculation of the HZ for
these systems.


\subsection{Known Planetary Systems}
\label{known}

From the 164 stars identified by \citet{mamajek2024}, we extracted the
30 stars that are currently known to host exoplanets
\citep{akeson2013}. The full list of planets for all of the included
systems, including their relevant stellar and planetary properties,
are provided in Table~\ref{tab:params}. The selection of relevant
parameters means those that are needed for the calculation of the HZ
and/or the dynamical simulations, described in
Section~\ref{dynamics}. Note that the HH~Peg (HD~206860) system was
not included in our analysis since the companion is a brown dwarf at
very wide separation from the host star, detected using the direct
imaging technique \citep{luhman2007a}. The sources for the
stellar/planetary parameters used in all cases are included in the
final column of Table~\ref{tab:params}, where the source selection is
based upon the most complete data source.

The stellar multiplicity of a planetary system can have a significant
dynamical effect on HZ planets
\citep{holman1999,kane2013a,simonetti2020}. Of the 30 known exoplanet
hosts in our sample, 9 of the stars are known to have stellar or brown
dwarf companions. The HD~3651 system has a brown dwarf with a
projected separation of 476~AU away the primary
\citep{luhman2007a}. The HD~9826 system contains a red dwarf star with
a projected separation of 750~AU from the primary \citep{mason2001c}.
HD~26965 is a triple system, whereby the B and C components are a
white dwarf and red dwarf star, respectively, and orbit each other at
a distance of $\sim$400~AU from the primary \citep{bond2017c}. The
HD~75732 system contains a red dwarf star with a projected separation
of 1100~AU from the primary \citep{marcy2002}. The HD~102365 includes
a red dwarf star with a projected separation of 211~AU from the
primary \citep{raghavan2010}. HD~115404A is in a binary orbit with a
red dwarf star that has projected separation of 289~AU
\citep{alonsofloriano2015a}. HD~147513 is in a binary orbit with a
white dwarf separated by $\sim$5300~AU \citep{portodemello1997a}. The
HD~190360 system includes a red dwarf star that is separated by at
least 1266~AU from the primary \citep{naef2003} and is probably at a
much larger distance \citep{feng2021}. The HD~209100 system contains a
brown dwarf binary pair that is separated from the primary by 1500~AU
\citep{scholz2003,luhman2007a,king2010b}. Given the wide separations
for all of these binary/multiple systems, and the considerable
uncertainties in their orbital parameters, the stellar or brown dwarf
companions were not included in our subsequent dynamical analysis.

There are several further important notes regarding the included
systems. The vast majority of the 70 planets are not known to transit
their host star, and so the planetary masses used are minimum masses
in most cases, the implications of which are discussed in
Section~\ref{discussion}. Exceptions to this include HD~39091~c
\citep{huang2018}, HD~75732~e
\citep{demory2011b,kane2011f,vonbraun2011b,winn2011b}, HD~136352 b, c,
and d \citep{kane2020c,delrez2021}, and HD~219134 b and c
\citep{gillon2017b}, all of which are known to transit their host
stars. There are also planets in our list for which orbital
inclination measurements have been made. For HD~39091, planet b has an
inclination of $\sim$50$^\circ$, resulting in a mutual inclination of
$\sim$40$^\circ$ with respect to planet c \citep{xuan2020b}, and we
assume planet d is coplanar with planet b. For HD~115404A, planet c
has an inclination of $\sim$25$^\circ$, and we assume planet b is
coplanar \citep{feng2022a}. For HD~140901, planet c has an inclination
of $\sim$170$^\circ$, and we assume planet b is coplanar
\citet{feng2022a}. For HD~160691, astrometry places an upper limit on
the inclination of $\sim$60$^\circ$ for the known planets
\citep{benedict2022b}. For HD~209100 b, \citet{feng2019b} provides an
inclination of $\sim$64$^\circ$, thus establishing the actual
planetary mass. Planet masses reported in the literature as Earth mass
units ($M_\oplus$) have been converted to Jupiter mass units ($M_J$)
in Table~\ref{tab:params} for consistency. For cases where the
eccentricity is reported as zero, the argument of periastron is set to
90$^\circ$, corresponding to the location of inferior
conjunction. Finally, note that the existence of the planet orbiting
HD~26965 \citep{ma2018} has been disputed from RV follow-up data
\citep{burrows2024}, but we elected to err on the side of caution and
included the planet in our list.

\begin{figure}
  \includegraphics[width=8.5cm]{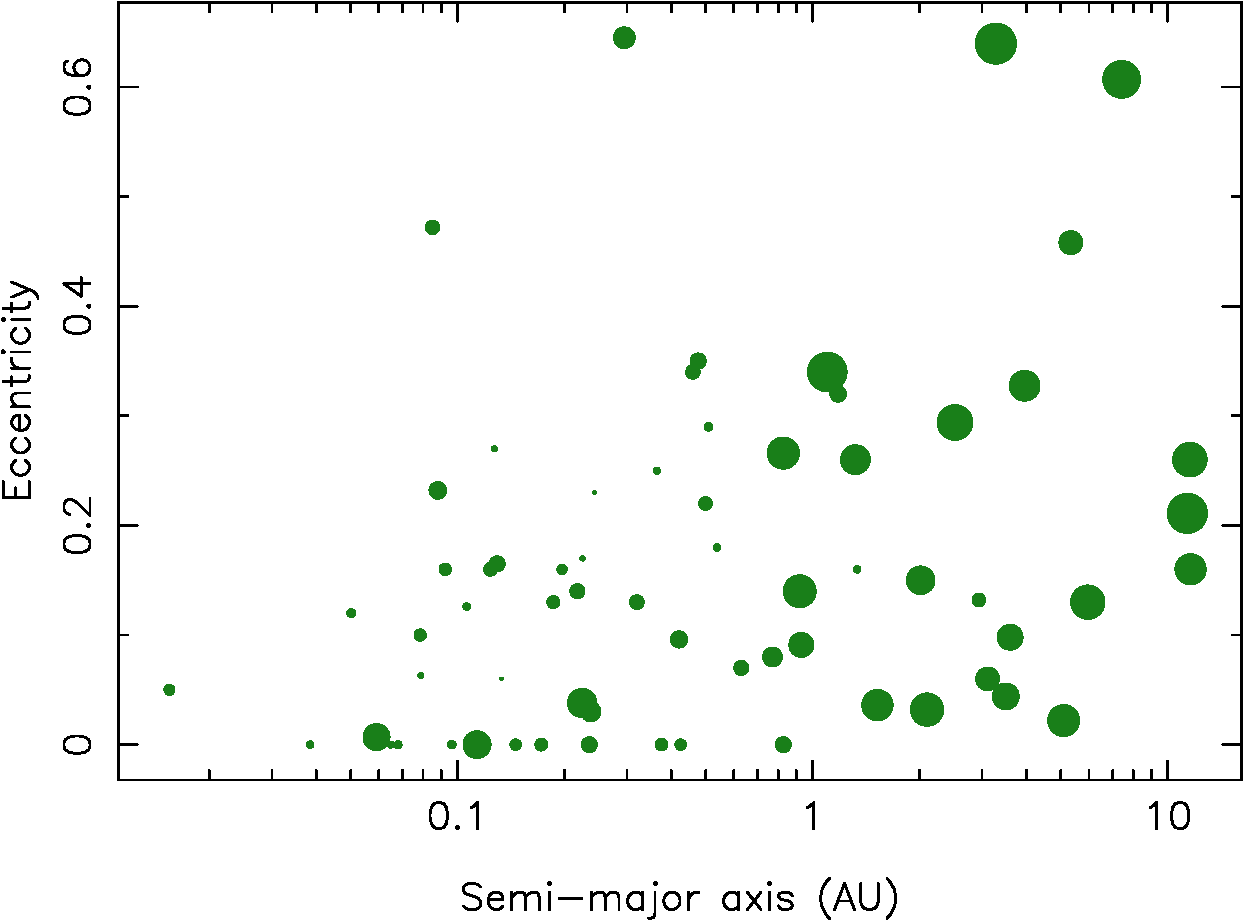}
  \caption{Distribution of semi-major axes and eccentricities for all
    planets included in our sample. The size of the plotted data are
    logarithmically proportional to the planet mass.}
  \label{fig:exo}
\end{figure}

The exoplanet population included in our sample is represented in
Figure~\ref{fig:exo}, showing the distributions of semi-major axes and
eccentricities. The ranges of these values within the sample are
0.015--11.6~AU and 0.0--0.64 for the semi-major axes and
eccentricities, respectively. The data point sizes are logarithmically
proportional to the planet mass, with a range of 0.0055~$M_J$ (tau
Ceti g) to 12.6~$M_J$ (pi Men b). There are several features to
consider regarding the parameter distribution shown in
Figure~\ref{fig:exo}. There is a clear trend toward higher mass
planets at larger separations. For example, the mean planet mass for
semi-major axes less than 1~AU is 0.19~$M_J$, whereas the mean planet
mass beyond 1~AU is 1.19~$M_J$. A significant factor in this planetary
mass trend is the observational bias inherent in the RV method,
whereby the RV amplitude decreases with increasing semi-major axis
\citep{cumming2004,otoole2009a}. A similar bias impacts the observed
eccentricity distribution, since such planets produce higher RV
amplitudes, albeit during a limited period of the orbital phase
\citep{kane2007a,shen2008c,zakamska2011,kane2012d}. Moreover, for
those long-period planets with insufficient phase coverage, the
relatively slow apastron passage may result in eccentricity
misclassifications
\citep{angladaescude2010a,hara2019,wittenmyer2019a}. Even so, the
combination of relatively high masses and eccentricities produces
numerous planetary architectures that exhibit significant dynamical
consequences within the HZ of their systems.

\startlongtable
\begin{deluxetable*}{lrrrlrrrrl}
  \tablecolumns{10}
  \tablewidth{0pc}
  \tablecaption{\label{tab:params} Stellar and planetary parameters
    for HWO known systems.}
  \tablehead{
    \colhead{Star} &
    \colhead{$NP^\dagger$} &
    \colhead{$M_\star$} &
    \colhead{$T_\mathrm{eff}$} &
    \colhead{Planet} &
    \colhead{$M_p^\ddagger$} &
    \colhead{$a$} &
    \colhead{$e$} &
    \colhead{$\omega$} &
    \colhead{Source} \\
    \colhead{} &
    \colhead{} &
    \colhead{($M_\odot$)} &
    \colhead{(K)} &
    \colhead{} &
    \colhead{($M_J$)} &
    \colhead{(AU)} &
    \colhead{} &
    \colhead{($^{\circ}$)} &
    \colhead{}
  }
  \startdata
HD 3651 (54 Piscium) & 1 & 0.799 & 5221 & b & 0.228 & 0.295 & 0.645 & 243 & \citet{wittenmyer2019b}\\
\hline
HD 9826 (ups And) & 3 & 1.29 & 6156 & b & 0.675 & 0.05914 & 0.0069 & 0 & \citet{rosenthal2021}\\
 &  &  &  & c & 1.965 & 0.8265 & 0.266 & 58.2 & \\
 &  &  &  & d & 4.1 & 2.517 & 0.294 & 73.8 & \\
\hline
HD 10647 (q1 Eridani) & 1 & 1.11 & 6218 & b & 0.94 & 2.015 & 0.15 & 212 & \citet{marmier2013}\\
\hline
HD 10700 (tau Ceti) & 4 & 0.78 & 5333 & g & 0.0055 & 0.133 & 0.06 & 395.34 & \citet{feng2017b}\\
 &  &  &  & h & 0.0058 & 0.243 & 0.23 & 7.45 & \\
 &  &  &  & e & 0.0124 & 0.538 & 0.18 & 22.35 & \\
 &  &  &  & f & 0.0124 & 1.334 & 0.16 & 119.75 & \\
\hline
HD 17051 (iota Hor) & 1 & 1.34 & 6167 & b & 2.27 & 0.92 & 0.14 & 309 & \citet{stassun2017}\\
\hline
HD 20794 (82 Eridani) & 4 & 0.813 & 5401 & b & 0.0089 & 0.127 & 0.27 & 386.17 & \citet{feng2017a}\\
 &  &  &  & c & 0.0079 & 0.225 & 0.17 & 31.51 & \\
 &  &  &  & d & 0.0111 & 0.364 & 0.25 & 252.67 & \\
 &  &  &  & e & 0.015 & 0.509 & 0.29 & 268.14 & \\
\hline
HD 22049 (eps Eri) & 1 & 0.81 & 5020 & b & 0.651 & 3.5 & 0.044 & 350 & \citet{rosenthal2021}\\
\hline
HD 26965 (40 Eri) & 1 & 0.78 & 5072 & b & 0.0267 & 0.219 & 0.04 & 148.969 & \citet{ma2018}\\
\hline
HD 33564 (kappa Cam) & 1 & 1.25 & 6250 & b & 9.1 & 1.1 & 0.34 & 205 & \citet{galland2005b}\\
\hline
HD 39091 (pi Men) & 3 & 1.07 & 5998 & c & 0.0114 & 0.06805 & 0 & 90 & \citet{hatzes2022}\\
 &  &  &  & d & 0.0421 & 0.499 & 0.22 & 323 & \\
 &  &  &  & b & 12.6 & 3.2826 & 0.6396 & 331.03 & \\
\hline
HD 69830 & 3 & 0.86 & 5385 & b & 0.0321 & 0.0785 & 0.1 & 340 & \citet{lovis2006}\\
 &  &  &  & c & 0.0371 & 0.186 & 0.13 & 221 & \\
 &  &  &  & d & 0.057 & 0.63 & 0.07 & 224 & \\
\hline
HD 75732 (55 Cancri) & 5 & 0.905 & 5172 & e & 0.0251 & 0.01544 & 0.05 & 86 & \citet{bourrier2018c}\\
 &  &  &  & b & 0.8036 & 0.1134 & 0 & 90 & \\
 &  &  &  & c & 0.1611 & 0.2373 & 0.03 & 2.4 & \\
 &  &  &  & f & 0.1503 & 0.7708 & 0.08 & 262.4 & \\
 &  &  &  & d & 3.12 & 5.957 & 0.13 & 290.9 & \\
\hline
HD 95128 (47 UMa) & 3 & 1.06 & 5872 & b & 2.53 & 2.1 & 0.032 & 334 & \citet{gregory2010}\\
 &  &  &  & c & 0.54 & 3.6 & 0.098 & 295 & \\
 &  &  &  & d & 1.64 & 11.6 & 0.16 & 110 & \\
\hline
HD 95735 & 2 & 0.3899 & 3712 & b & 0.0085 & 0.07879 & 0.063 & 330 & \citet{hurt2022}\\
 &  &  &  & c & 0.0428 & 2.94 & 0.132 & 63 & \\
\hline
HD 102365 & 1 & 0.85 & 5630 & b & 0.0503 & 0.46 & 0.34 & 105 & \citet{tinney2011a}\\
\hline
HD 114613 & 1 & 1.27 & 5641 & b & 0.357 & 5.34 & 0.458 & 196 & \citet{luhn2019}\\
\hline
HD 115404A & 2 & 0.83 & 5019 & b & 0.097 & 0.088 & 0.232 & 259 & \citet{feng2022a}\\
 &  &  &  & c & 10.319 & 11.364 & 0.211 & 312 & \\
\hline
HD 115617 (61 Vir) & 3 & 0.942 & 5577 & b & 0.016 & 0.050201 & 0.12 & 105 & \citet{vogt2010a}\\
 &  &  &  & c & 0.057 & 0.2175 & 0.14 & 341 & \\
 &  &  &  & d & 0.072 & 0.476 & 0.35 & 314 & \\
\hline
HD 136352 (nu2 Lupi) & 3 & 0.87 & 5664 & b & 0.0149 & 0.0964 & 0 & 90 & \citet{delrez2021}\\
 &  &  &  & c & 0.0354 & 0.1721 & 0 & 90 & \\
 &  &  &  & d & 0.0278 & 0.425 & 0 & 90 & \\
\hline
HD 141004 (lambda Ser) & 1 & 1.05 & 5885 & b & 0.0428 & 0.1238 & 0.16 & 19.48 & \citet{rosenthal2021}\\
\hline
HD 140901 & 2 & 0.99 & 5586 & b & 0.0503 & 0.085 & 0.472 & 108 & \citet{feng2022a}\\
 &  &  &  & c & 6.284 & 7.421 & 0.607 & 298 & \\
\hline
HD 143761 (rho CrB) & 4 & 0.95 & 5817 & e & 0.0119 & 0.1061 & 0.126 & 359.4 & \citet{brewer2023}\\
 &  &  &  & b & 1.093 & 0.2245 & 0.038 & 269.64 & \\
 &  &  &  & c & 0.0887 & 0.4206 & 0.096 & 9.7 & \\
 &  &  &  & d & 0.068 & 0.827 & 0 & 90 & \\
\hline
HD 147513 (62 G. Sco) & 1 & 1.11 & 5883 & b & 1.21 & 1.32 & 0.26 & 282 & \citet{mayor2004}\\
\hline
HD 160691 (mu Ara) & 4 & 1.13 & 5773 & d & 0.033 & 0.0923 & 0.16 & 197 & \citet{benedict2022b}\\
 &  &  &  & e & 0.439 & 0.9296 & 0.091 & 193 & \\
 &  &  &  & b & 1.665 & 1.5224 & 0.036 & 39 & \\
 &  &  &  & c & 1.873 & 5.0937 & 0.022 & 84 & \\
\hline
HD 189567 & 2 & 0.83 & 5726 & b & 0.0267 & 0.111 & 0 & 90 & \citet{unger2021}\\
 &  &  &  & c & 0.022 & 0.197 & 0.16 & 219 & \\
\hline
HD 190360 & 2 & 0.99 & 5537 & c & 0.0677 & 0.1294 & 0.165 & 322 & \citet{rosenthal2021}\\
 &  &  &  & b & 1.492 & 3.955 & 0.3274 & 13.9 & \\
\hline
HD 192310 & 2 & 0.8 & 5166 & b & 0.0532 & 0.32 & 0.13 & 173 & \citet{pepe2011}\\
 &  &  &  & c & 0.076 & 1.18 & 0.32 & 110 & \\
\hline
HD 209100 (eps Indi A) & 1 & 0.754 & 4611 & b & 3.25 & 11.55 & 0.26 & 77.83 & \citet{feng2019b}\\
\hline
HD 217987 & 2 & 0.489 & 3688 & b & 0.0132 & 0.068 & 0 & 90 & \citet{jeffers2020}\\
 &  &  &  & c & 0.0239 & 0.12 & 0 & 90 & \\
\hline
HD 219134 & 6 & 0.794 & 4913 & b & 0.012 & 0.038474 & 0 & 90 & \citet{vogt2015}\\
 &  &  &  & c & 0.011 & 0.064816 & 0 & 90 & \\
 &  &  &  & f & 0.028 & 0.14574 & 0 & 90 & \\
 &  &  &  & d & 0.067 & 0.23508 & 0 & 90 & \\
 &  &  &  & g & 0.0346 & 0.3753 & 0 & 90 & \\
 &  &  &  & h & 0.34 & 3.11 & 0.06 & 215 & \\
  \enddata
  \tablenotetext{\dagger}{Number of planets.}
  \tablenotetext{\ddagger}{Minimum mass, except for 55~Cancri~e and the HD~136352 planets.}
\end{deluxetable*}


\subsection{System Habitable Zones}
\label{hz}

\begin{figure*}
    \begin{center}
        \begin{tabular}{cc}
            \includegraphics[width=8.0cm]{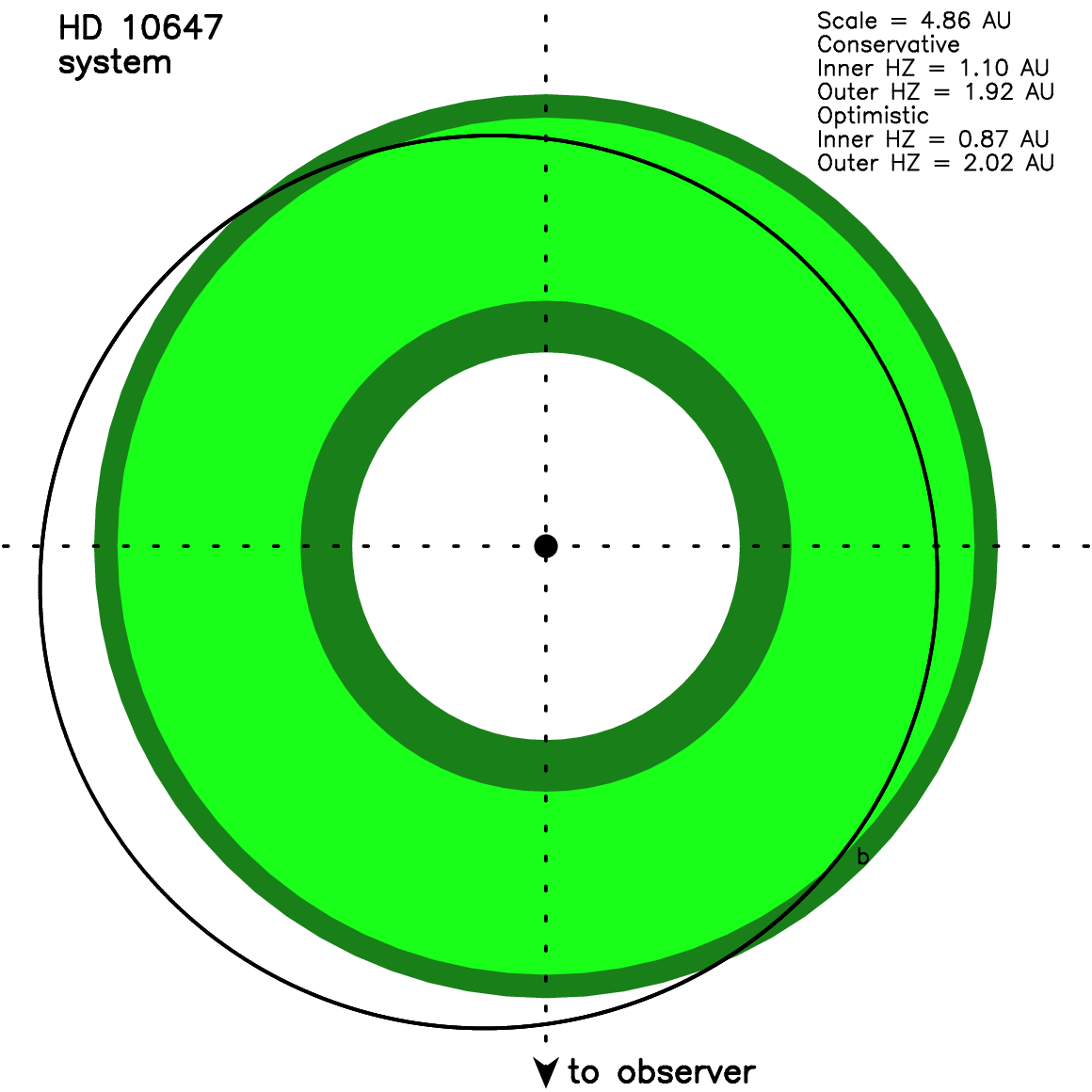} &
            \includegraphics[width=8.0cm]{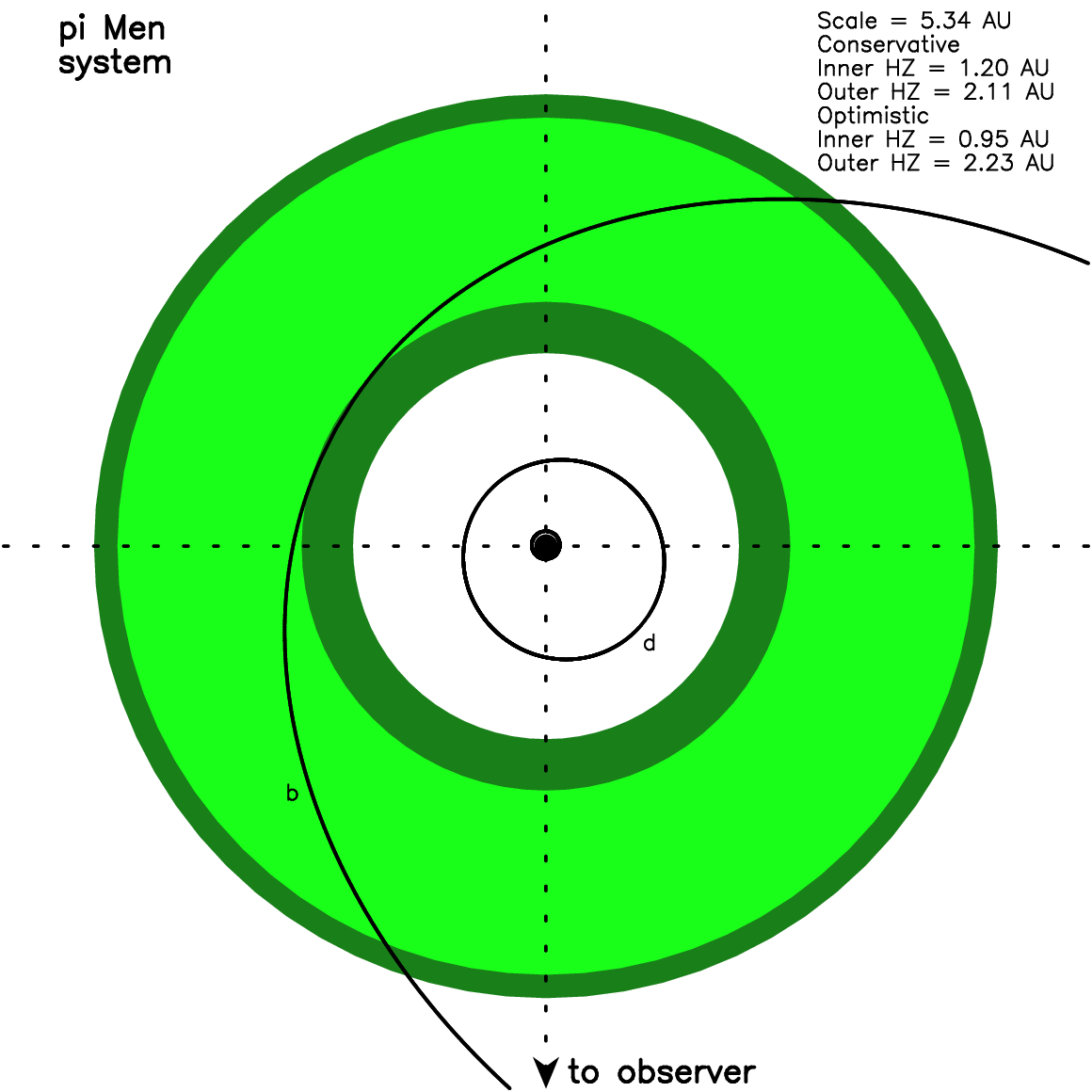} \\
            \includegraphics[width=8.0cm]{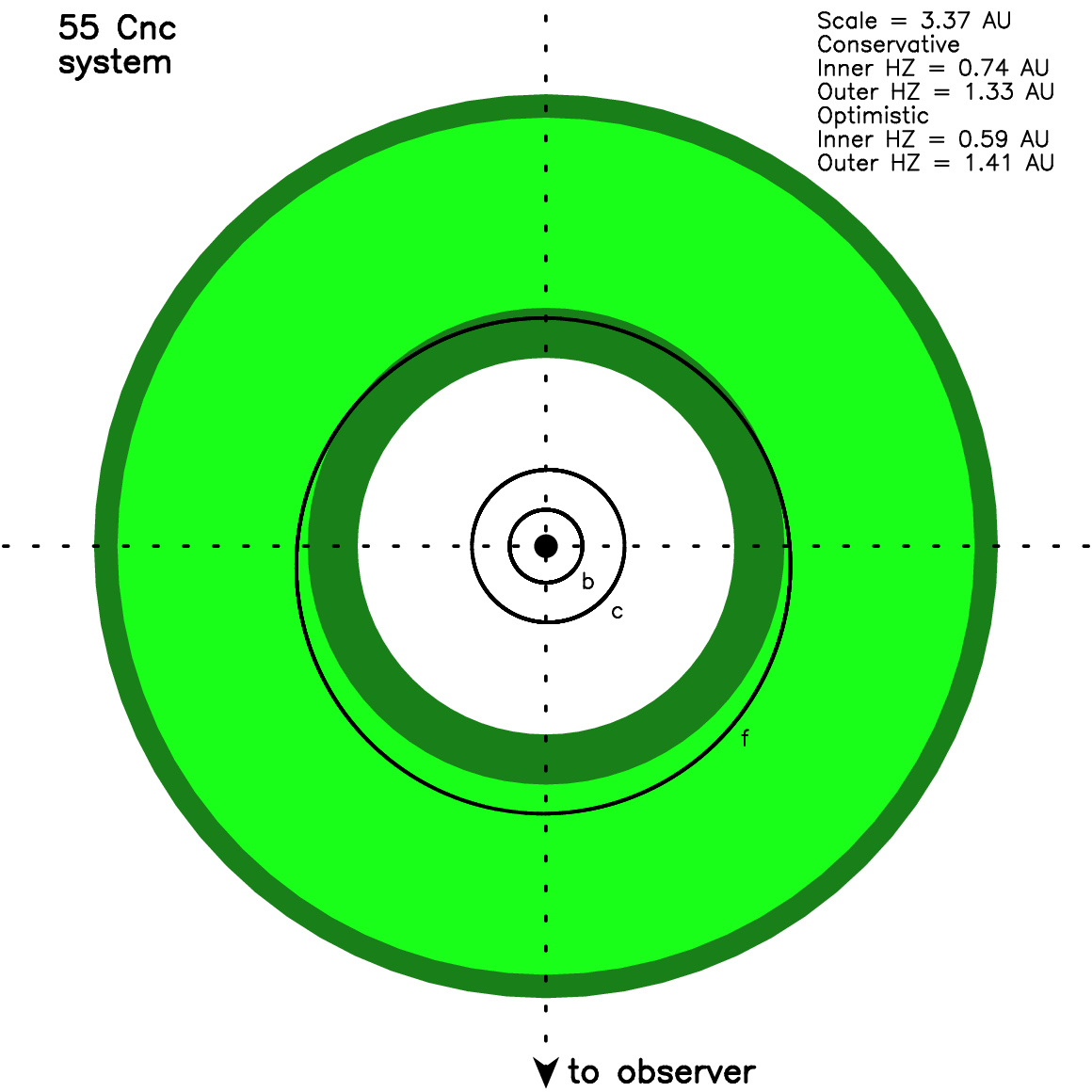} &
            \includegraphics[width=8.0cm]{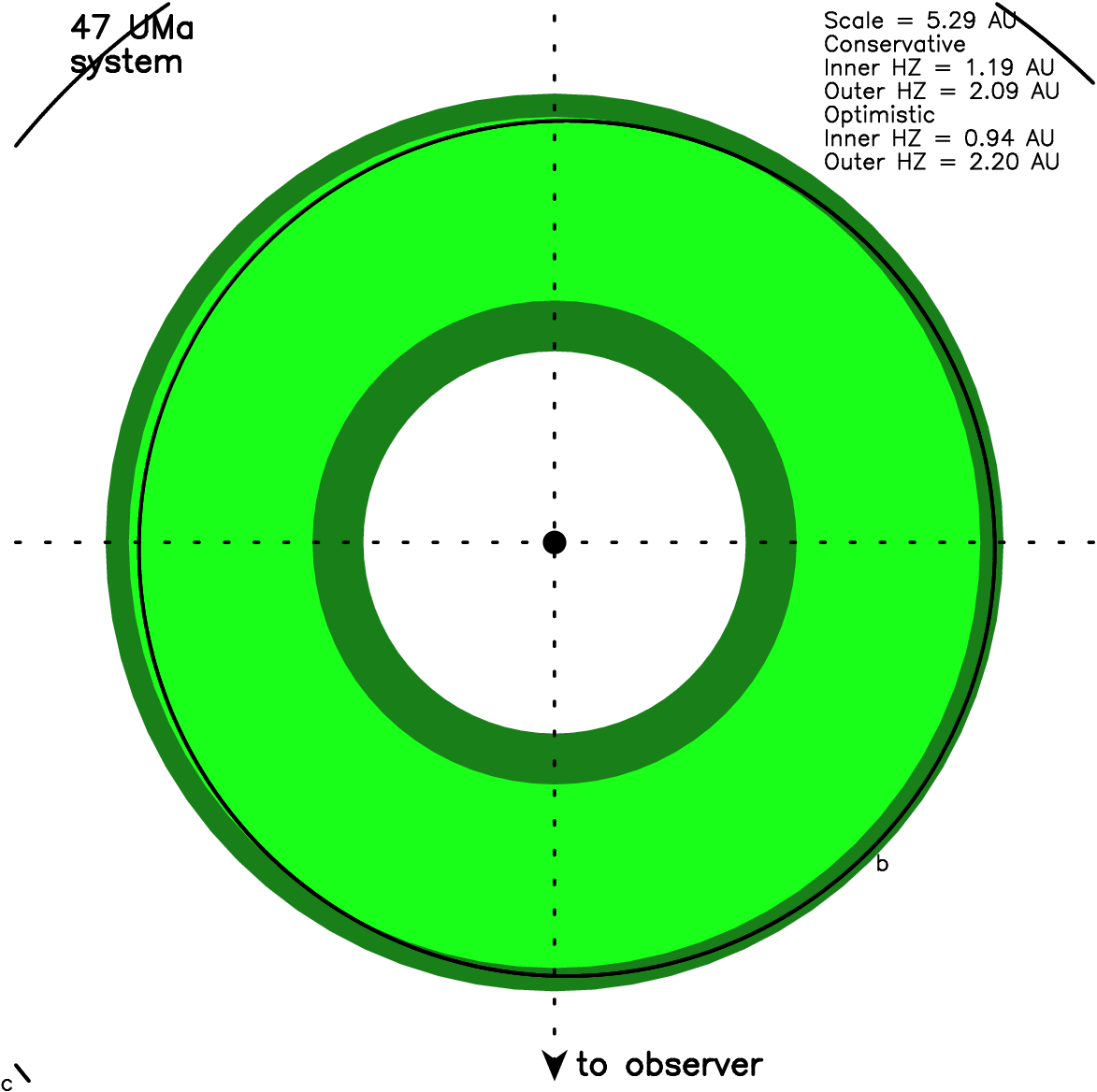}
        \end{tabular}
    \end{center}
  \caption{HZ and planetary orbits for four of the systems in our
    sample: HD~10647 (top-left), HD~39091 (pi Men; top-right),
    HD~75732 (55 Cancri; bottom-left), HD~95128 (47 Uma;
    bottom-right). The orbits are labeled by planet designation. The
    extent of the HZ is shown in green, where light green and dark
    green indicate the CHZ and OHZ, respectively.}
  \label{fig:hz1}
\end{figure*}

\begin{figure*}
    \begin{center}
        \begin{tabular}{cc}
            \includegraphics[width=8.0cm]{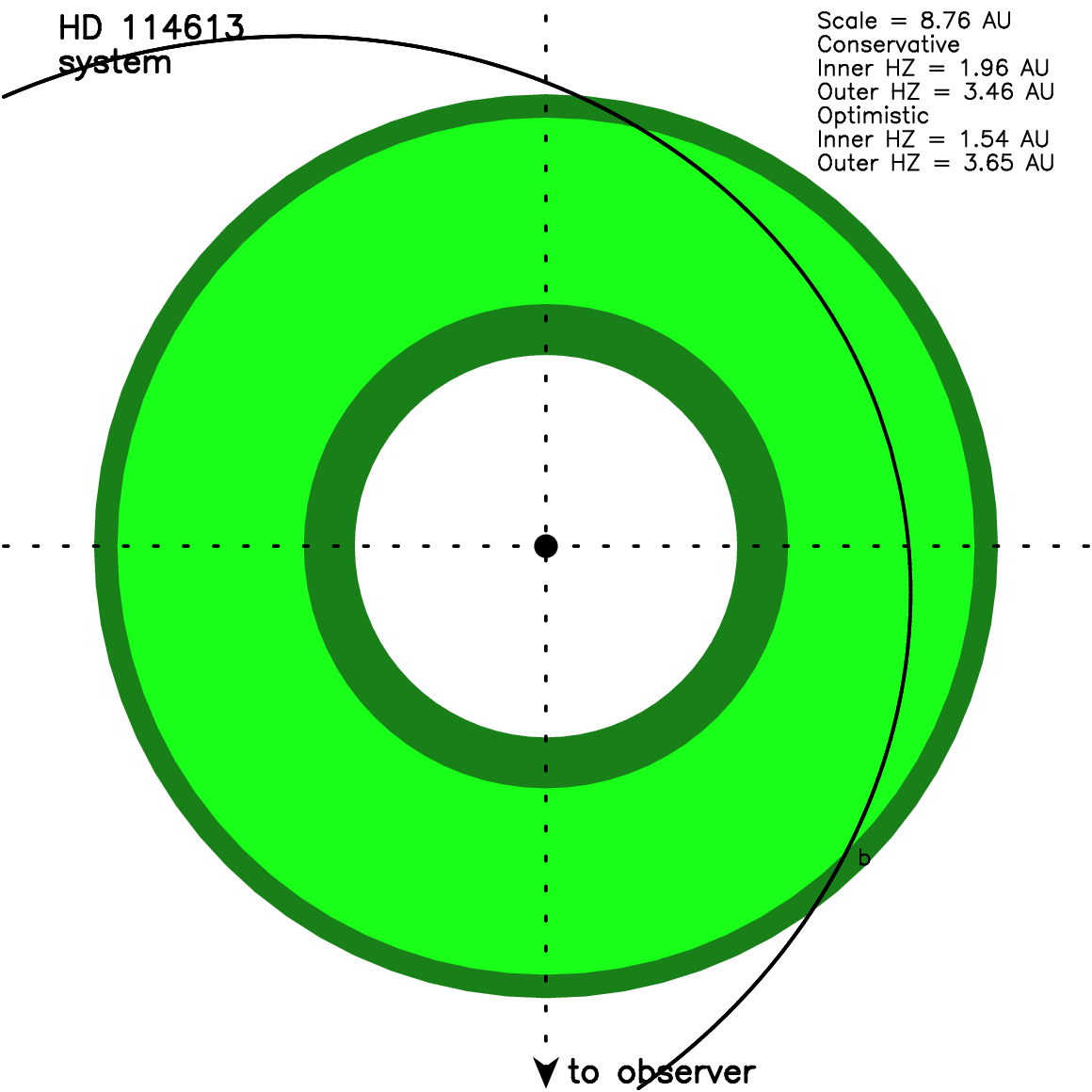} &
            \includegraphics[width=8.0cm]{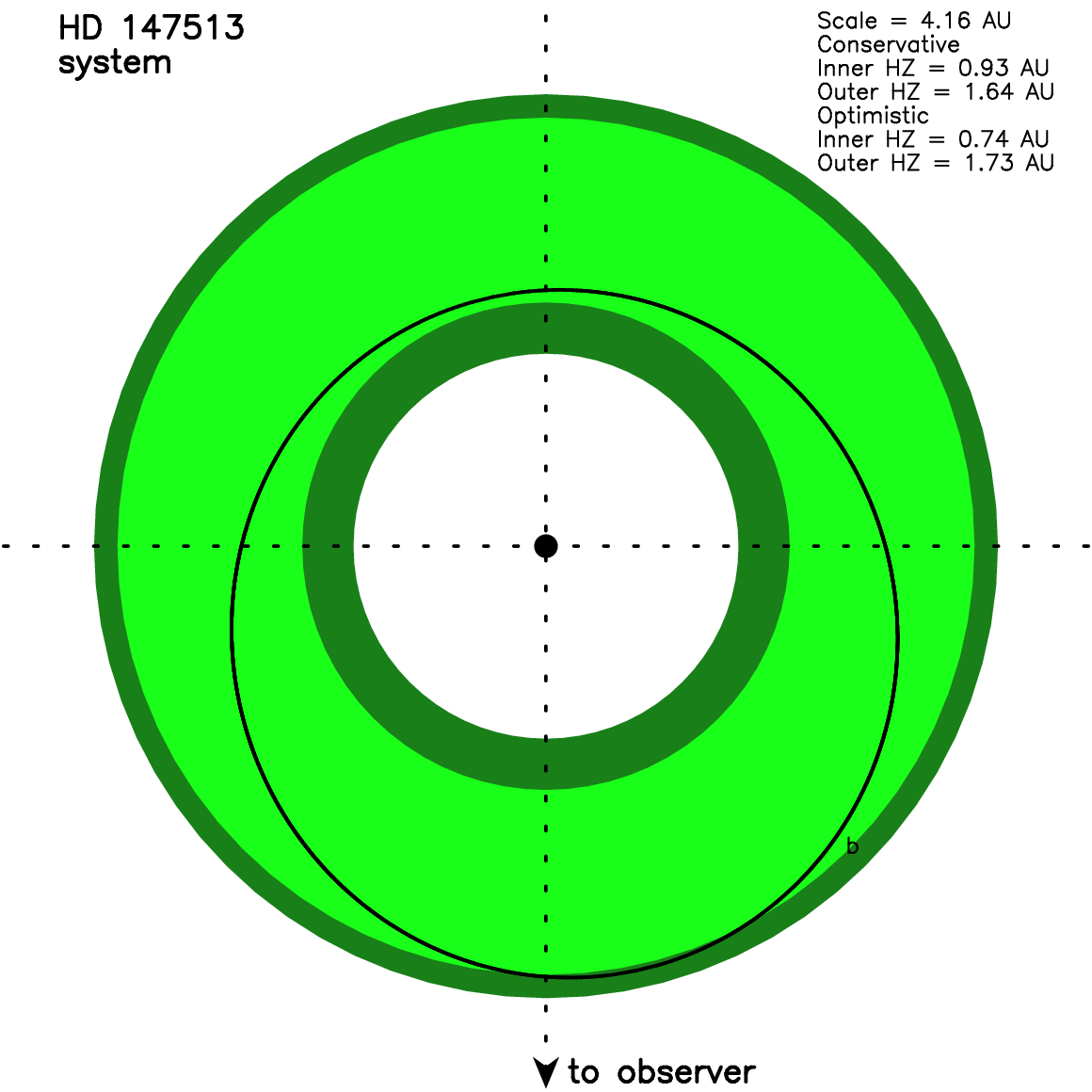} \\
            \includegraphics[width=8.0cm]{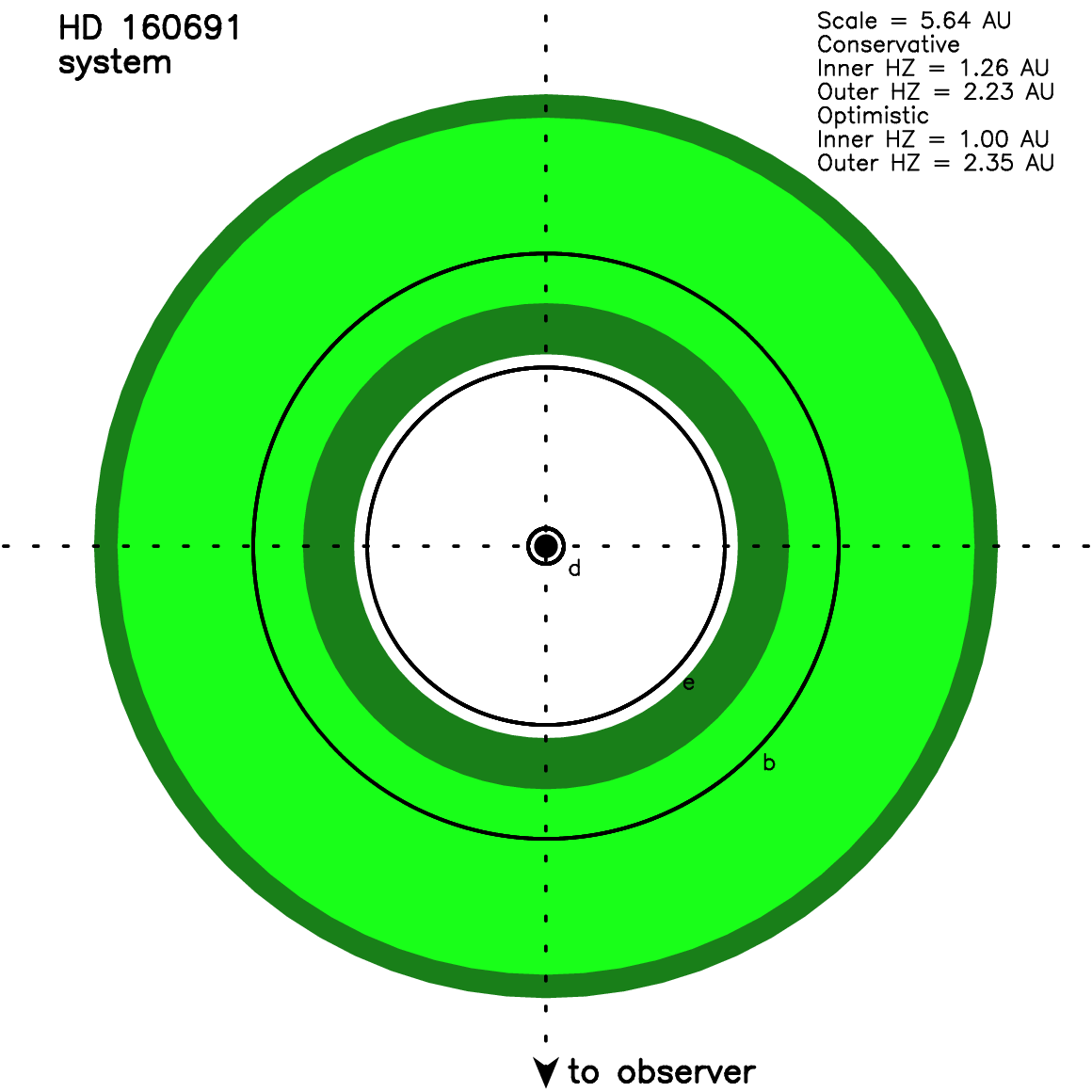} &
            \includegraphics[width=8.0cm]{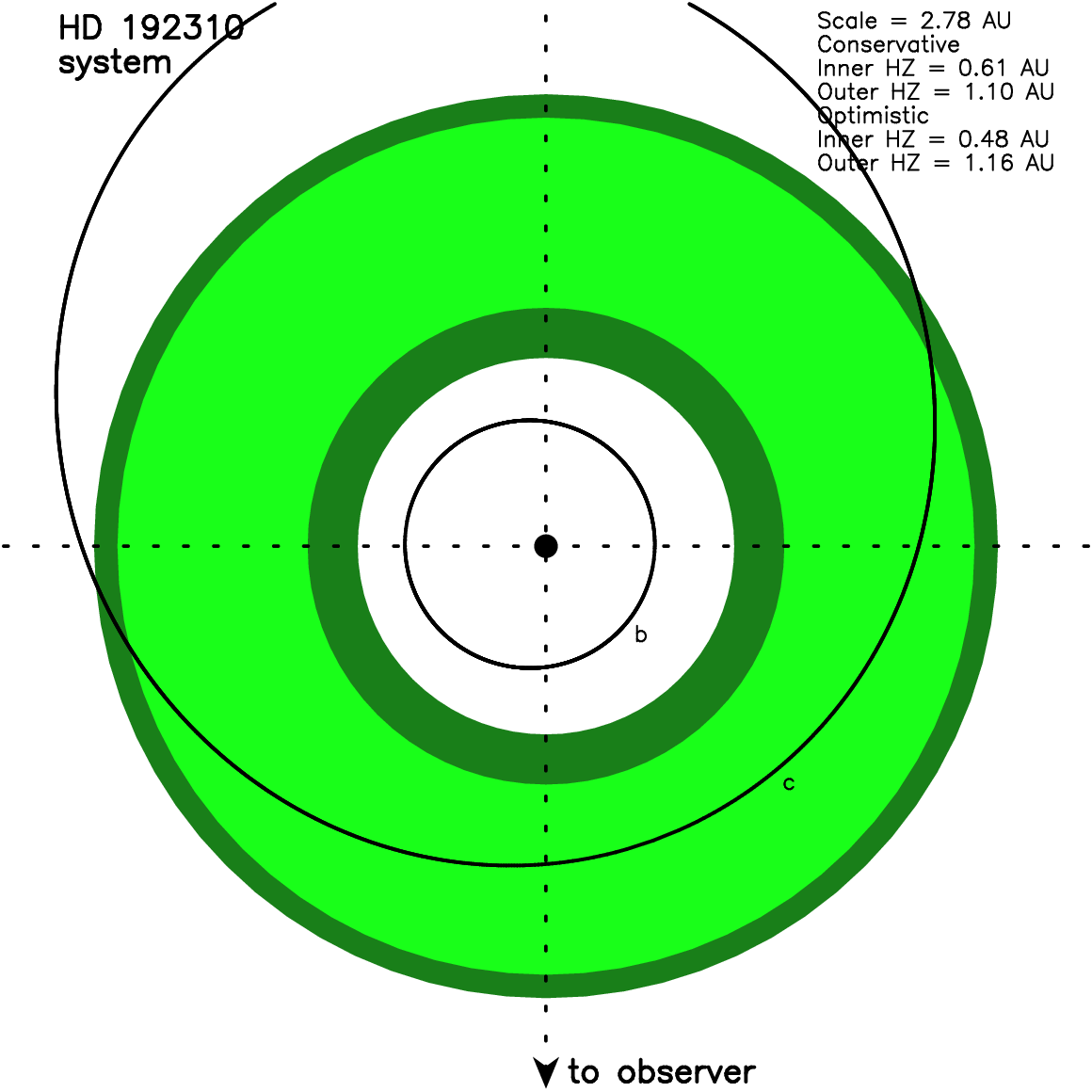}
        \end{tabular}
    \end{center}
  \caption{HZ and planetary orbits for four of the systems in our
    sample: HD~114613 (top-left), HD~147513 (top-right), HD~160691
    (bottom-left), HD~192310 (bottom-right). The orbits are labeled by
    planet designation. The extent of the HZ is shown in green, where
    light green and dark green indicate the CHZ and OHZ,
    respectively.}
  \label{fig:hz2}
\end{figure*}

The extent of the HZ for each system was calculated using the stellar
parameters from the sources provided in
Table~\ref{tab:params}. Specifically, we adopt the methodology
provided by \citet{kopparapu2013a,kopparapu2014}, in which the
boundaries of the HZ are derived from 1D Earth-based climate models
that calculate the radiative balance for which surface liquid water is
retained. This approach divides the HZ into two primary regions: the
conservative HZ (CHZ) and the optimistic HZ (OHZ). The CHZ is bounded
by the runaway greenhouse criteria at the inner edge, and the maximum
CO$_2$ greenhouse criteria at the outer edge \citep{kane2016c}. The
OHZ is an empirical extension to the CHZ based upon assumptions
regarding retention of surface liquid water in the Venusian and
Martian evolutionary histories
\citep{baker2001,kane2014e,way2016,orosei2018,kane2019d,kane2024b}. The
CHZ and OHZ boundaries for each system are provided in
Table~\ref{tab:results}.

Figure~\ref{fig:hz1} and Figure~\ref{fig:hz2} show top-down views for
8 of the planetary systems, that demonstrate the diversity of
architectures in our known exoplanet host sample. The extent of the
CHZ and OHZ are shown in light green and dark green, respectively. In
most of these cases, the planetary orbit that is present within the HZ
is a giant planet with significant perturbation potential. The cases
where eccentric orbits lie within and/or cross the HZ, such as
HD~39091 (pi Men), HD~114613, HD~147513, and HD~192310, are
particularly devastating to orbital stability in that region due to
angular momentum transfer that excites other planetary orbits into
high eccentricity regimes \citep{kane2014b}. The others cases shown
exhibit planetary orbits that are more circular, and can often allow
regions of orbital stability at the edges of the HZ. These scenarios
may be tested with dynamical simulations that explore particle
injection throughout the HZ region.


\section{Orbital Dynamics}
\label{dynamics}


\subsection{Simulation Design}
\label{sim}

We conducted an exhaustive series of dynamical simulations to test the
viability of terrestrial planets orbiting within the HZ of the target
systems. The simulations were conducted using the REBOUND N-body
integrator package \citep{rein2012a} that applies the symplectic
integrator WHFast \citep{rein2015c}. Our simulations explored the
effect and orbital evolution of an additional Earth-mass planet in a
circular orbit that is coplanar with the other planets within each
system, where 10 equally spaced mean anomalies were chosen for the
injected planet. For the case of HD~39091 (pi~Men), where there is a
substantial mutual inclination between the planets, we injected the
additional planet in an orbit coplanar with planet b, since that is
the planet that dominates the gravitational influence on the HZ for
that system (see Section~\ref{known} and Figure~\ref{fig:hz1}). The
semi-major axes explored spanned 20\% less than the inner edge of the
OHZ to 20\% larger than the outer edge of the OHZ (see
Section~\ref{hz}). This region within each system was divided into
1000 equally spaced semi-major axis starting locations for the
injected planet to ensure sufficiently dynamical sampling of the
HZ. Each of these semi-major axis and mean anomaly starting locations
were integrated for $10^7$ years, resulting in 10,000 simulations
performed for each system. The time step for each integration was
selected based upon the orbital period for the innermost body within
the system divided by 20, where the innermost body is either a known
planet or the injected planet.

At each of the simulation starting locations, our simulations provided
the percentage of the total simulation duration for which the planet
was able to survive, where non-survival means that the injected planet
was captured by the gravitational well of the host star or ejected
from the system. From these results, we calculated the dynamically
viable HZ (DVHZ), which is the percentage of the OHZ locations for
which the injected planet survived the full $10^7$ year simulation.


\subsection{Habitable Zone Stability Results}
\label{results}

The methodology described in Section~\ref{sim} was applied to each of
the systems shown in Table~\ref{tab:params}. The time required to
complete the suite of simulations for each system varied enormously,
depending upon the number of planets in the system, their locations,
and the extent of the instability induced within the HZ. For example,
the 55 Cancri system (bottom-left system depicted in
Figure~\ref{fig:hz1}) contains 5 known planets, including one with a
0.74~day orbit, resulting in the need for high time resolution within
the dynamical simulation. By contrast, systems with a single,
eccentric giant planet with high scattering potential (such as the
HD~147513 system) concluded the simulation suite relatively quickly.

\begin{figure*}
  \begin{center}
    \includegraphics[width=16.0cm,height=5.2cm]{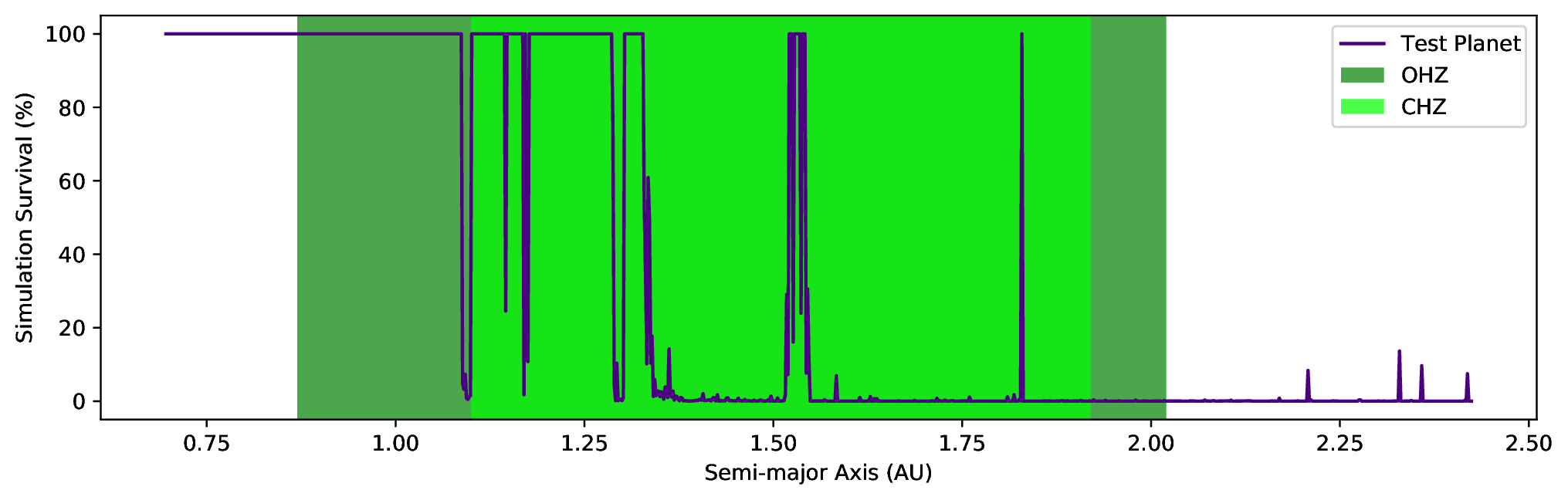} \\
    \includegraphics[width=16.0cm,height=5.2cm]{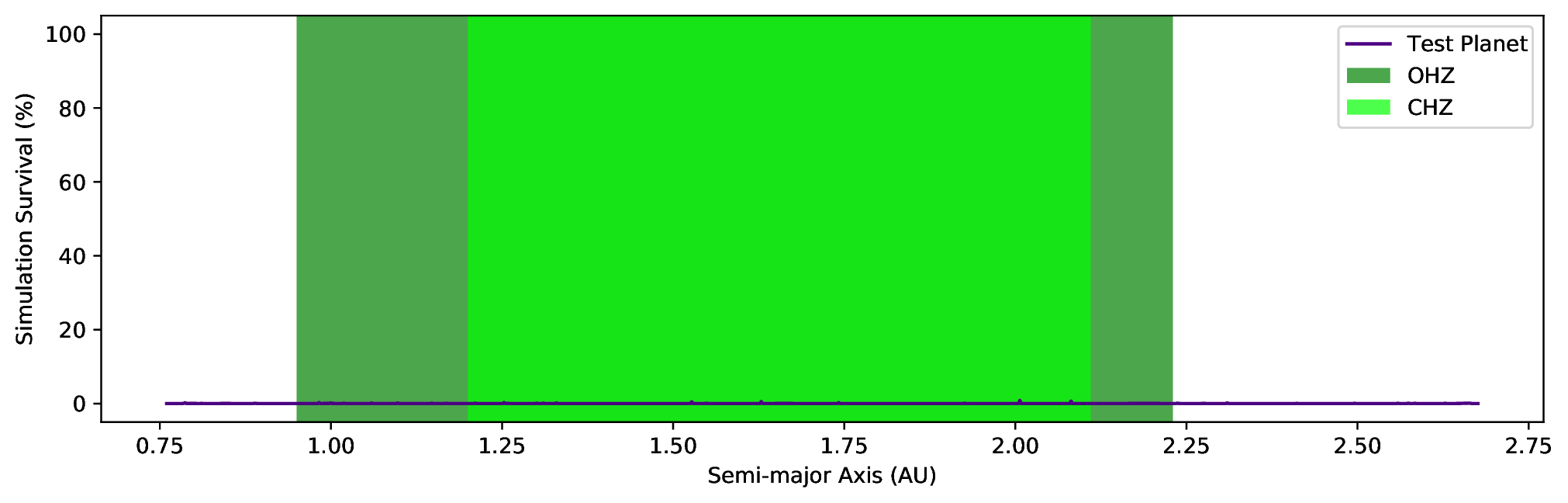} \\
    \includegraphics[width=16.0cm,height=5.2cm]{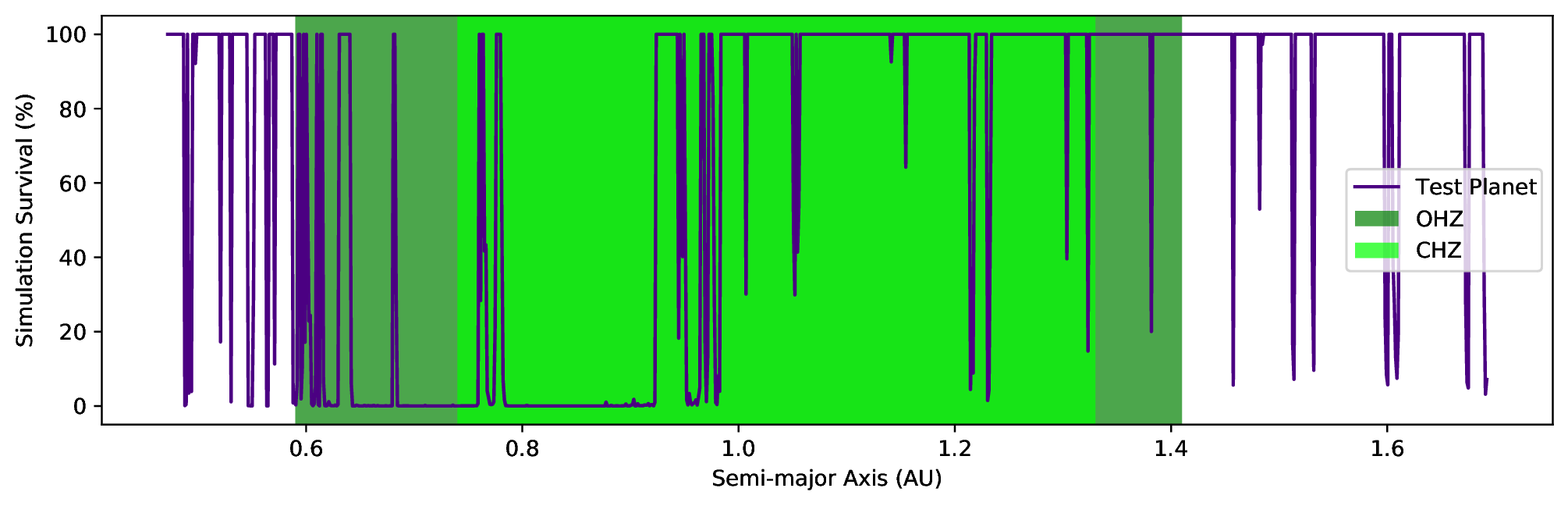} \\
    \includegraphics[width=16.0cm,height=5.2cm]{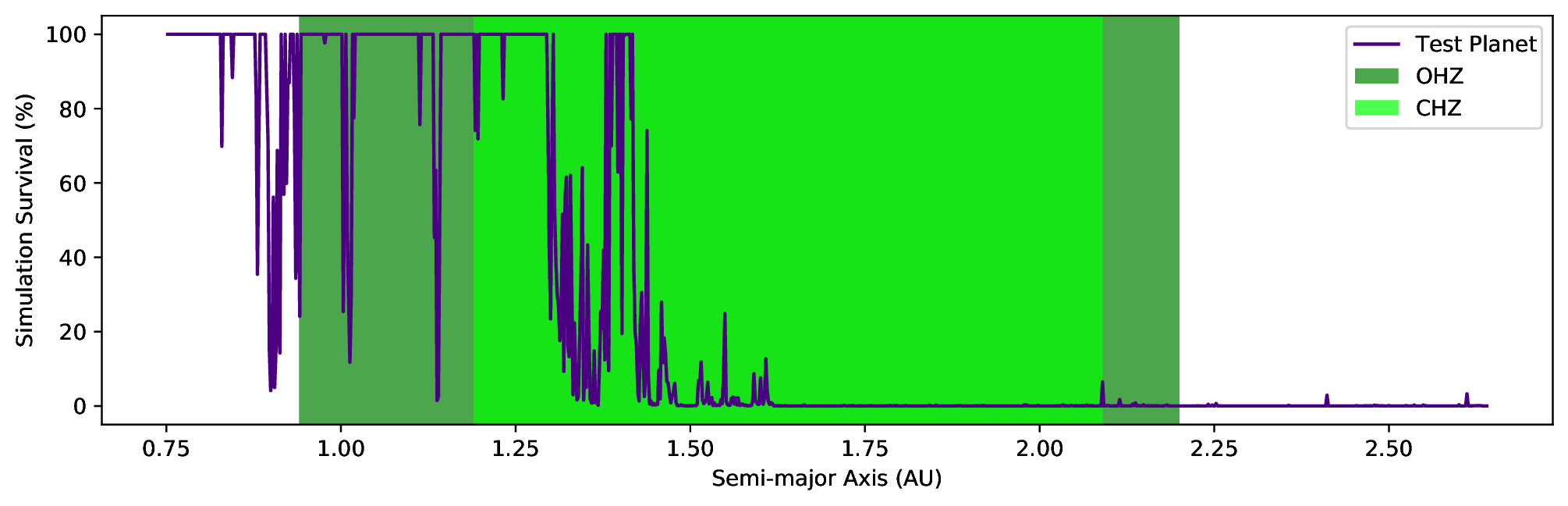}
    \end{center}
  \caption{Percentage of the simulation that the injected Earth-mass
    planet survived as a function of the initial semi-major axis, where the CHZ is
    shown in light green and the OHZ is shown in dark green. Shown,
    from top to bottom, are the results for HD~10647, HD~39091 (pi
    Men), HD~75732 (55 Cancri), and HD~95128.}
  \label{fig:stab1}
\end{figure*}

\begin{figure*}
  \begin{center}
    \includegraphics[width=16.0cm,height=5.2cm]{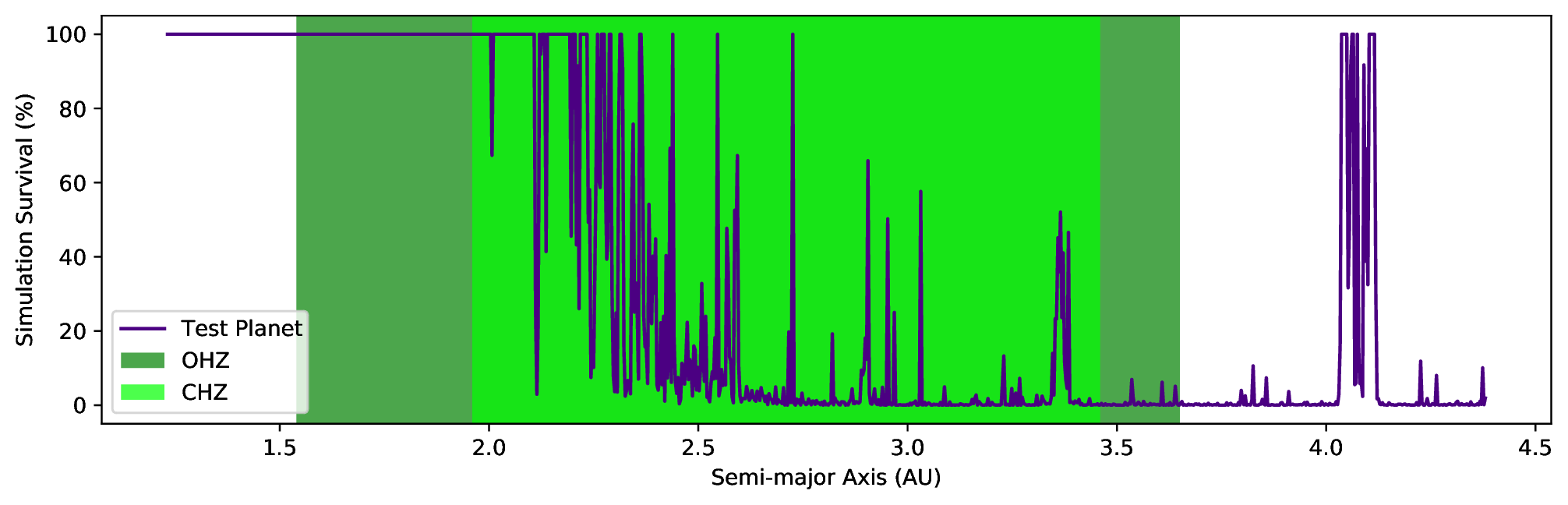} \\
    \includegraphics[width=16.0cm,height=5.2cm]{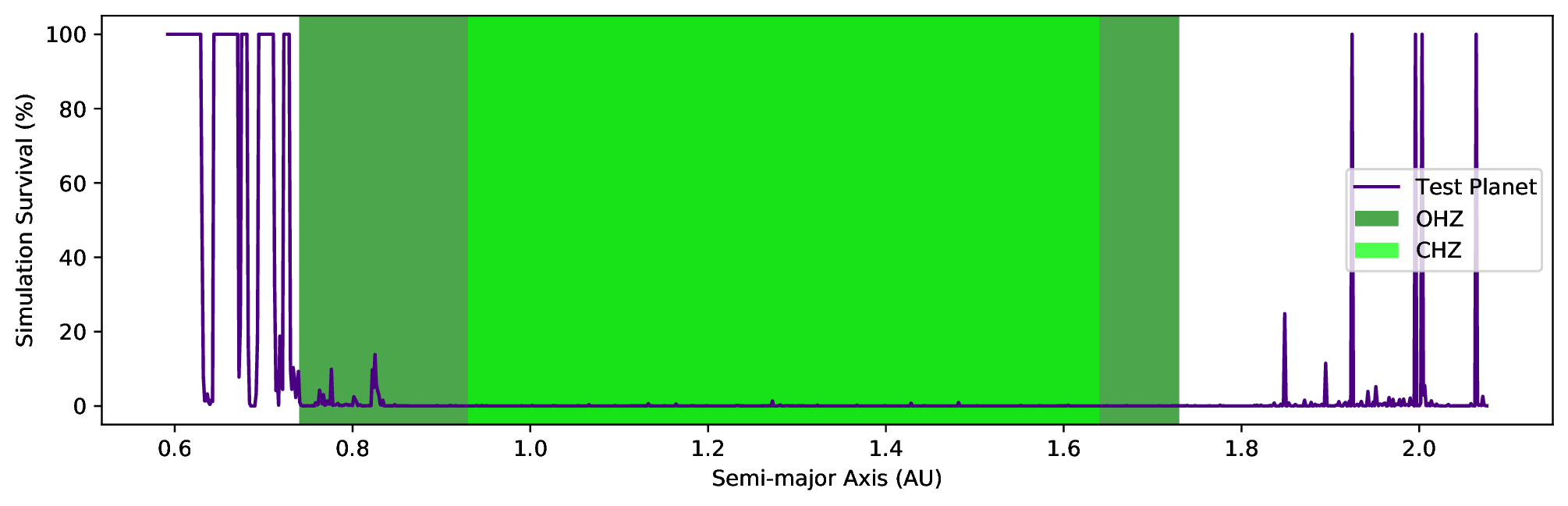} \\
    \includegraphics[width=16.0cm,height=5.2cm]{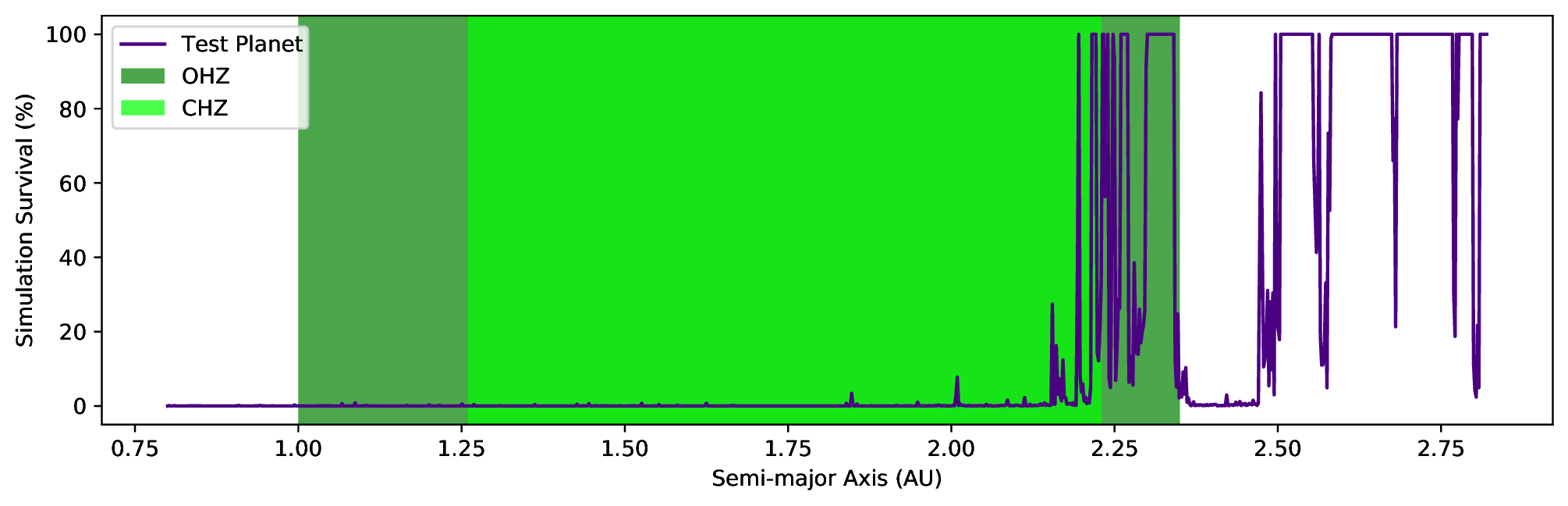} \\
    \includegraphics[width=16.0cm,height=5.2cm]{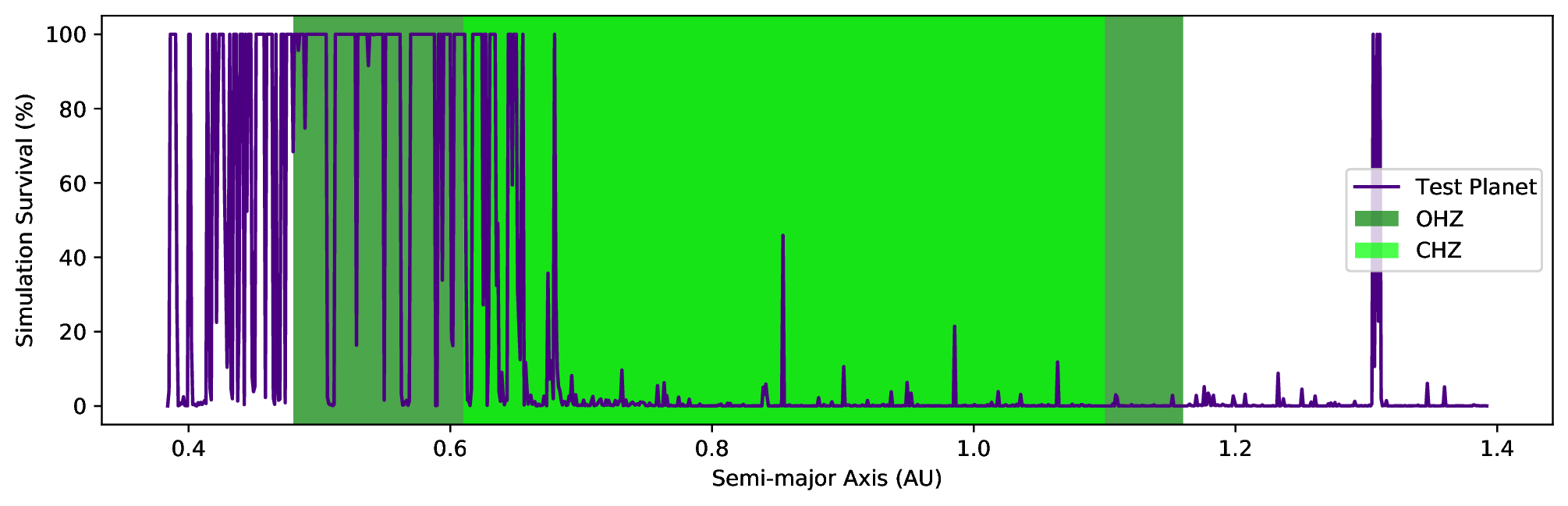}
    \end{center}
  \caption{Percentage of the simulation that the injected Earth-mass
    planet survived as a function of the initial semi-major axis, where the CHZ is
    shown in light green and the OHZ is shown in dark green. Shown,
    from top to bottom, are the results for HD~114613, HD~147513,
    HD~160691, and HD~192310.}
  \label{fig:stab2}
\end{figure*}

The simulation results for 8 of the systems are represented in
Figure~\ref{fig:stab1} and Figure~\ref{fig:stab2}, corresponding to
the same systems shown in Figure~\ref{fig:hz1} and
Figure~\ref{fig:hz2}, respectively. The plots provide the percentage
of the full $10^7$ year simulation for which the injected terrestrial
planet was able to maintain a stable orbit for each semi-major axis
location. As for Figure~\ref{fig:hz1} and Figure~\ref{fig:hz2}, the
CHZ and OHZ are shown as light green and dark green shaded regions,
respectively. In all of the 8 cases shown, there are significant areas
of instability within the HZ resulting from the known planets in the
system. Most notable is the case of pi Men (second panel;
Figure~\ref{fig:stab1}) in which the entire region explored is
unstable due to the incursion of the highly eccentric b planet into
the HZ. The HD~147513 system (second panel; Figure~\ref{fig:stab2}) is
similarly affected, although stable regions exist just outside of the
HZ. The remaining examples shown have varying amounts of stable
locations within the HZ, including regions of stability/instability
resulting from a combination of mean motion resonance (MMR) effects
and the numerical randomness intrinsic to the simulations
\citep{raymond2008b,petrovich2013,hadden2019b}. For example, the
results for HD~10647 (first panel; Figure~\ref{fig:stab1}) exhibits
numerous stability islands induced by the single giant planet in the
system. The 3:2 MMR location at 1.53~AU creates an island of stability
whereby a planet located there has the potential to avoid
gravitational perturbing interactions with the known giant planet,
similar to the 2:3 MMR of Neptune with Pluto \citep{williams1971e}. Of
particular note is the 5:2 MMR location at 1.09~AU that produces a
region of instability that coincides with the inner boundaries of the
OHZ and CHZ. Multi-planet systems can exhibit significantly
complicated stability patterns due to the multitude of MMR locations
cascading throughout the system, as shown in the case of HD~75732 (55
Cancri; third panel; Figure~\ref{fig:stab1}).

\begin{deluxetable}{lrrrrr}
  \tablecolumns{6}
  \tablewidth{0pc}
  \tablecaption{\label{tab:results} HZ boundaries and DVHZ results.}
  \tablehead{
    \colhead{Star} &
    \colhead{OHZ$_\mathrm{in}$} &
    \colhead{CHZ$_\mathrm{in}$} &
    \colhead{CHZ$_\mathrm{out}$} &
    \colhead{OHZ$_\mathrm{out}$} &
    \colhead{DVHZ} \\
    \colhead{} &
    \colhead{(AU)} &
    \colhead{(AU)} &
    \colhead{(AU)} &
    \colhead{(AU)} &
    \colhead{(\%)}
  }
  \startdata
HD 3651    & 0.56 & 0.71 & 1.27 & 1.34 & 92.41 \\
HD 9826    & 1.35 & 1.71 & 3.00 & 3.16 & 0.00 \\
HD 10647   & 0.87 & 1.10 & 1.92 & 2.02 & 38.65 \\
HD 10700   & 0.55 & 0.69 & 1.24 & 1.31 & 20.75 \\
HD 17051   & 0.94 & 1.20 & 2.09 & 2.20 & 55.02 \\
HD 20794   & 0.72 & 0.91 & 1.63 & 1.72 & 0.00 \\
HD 22049   & 0.45 & 0.57 & 1.03 & 1.09 & 100.00 \\
HD 26965   & 0.52 & 0.66 & 1.20 & 1.26 & 100.00 \\
HD 33564   & 1.58 & 2.00 & 3.49 & 3.68 & 61.50 \\
HD 39091   & 0.95 & 1.20 & 2.11 & 2.23 & 0.00 \\
HD 69830   & 0.60 & 0.75 & 1.35 & 1.42 & 84.48 \\
HD 75732   & 0.59 & 0.74 & 1.33 & 1.41 & 56.99 \\
HD 95128   & 0.94 & 1.19 & 2.09 & 2.20 & 27.59 \\
HD 95735   & 0.13 & 0.16 & 0.30 & 0.32 & 100.00 \\
HD 102365  & 0.57 & 0.72 & 1.28 & 1.35 & 78.21 \\
HD 114613  & 1.54 & 1.96 & 3.46 & 3.65 & 33.43 \\
HD 115404A & 0.43 & 0.54 & 0.99 & 1.04 & 100.00 \\
HD 115617  & 0.68 & 0.86 & 1.53 & 1.61 & 44.93 \\
HD 136352  & 0.77 & 0.97 & 1.72 & 1.82 & 100.00 \\
HD 140901  & 0.69 & 0.87 & 1.54 & 1.62 & 89.06 \\
HD 141004  & 1.08 & 1.36 & 2.40 & 2.53 & 100.00 \\
HD 143761  & 1.01 & 1.28 & 2.25 & 2.38 & 88.17 \\
HD 147513  & 0.74 & 0.93 & 1.64 & 1.73 & 0.00 \\
HD 160691  & 1.00 & 1.26 & 2.23 & 2.35 & 5.39 \\
HD 189567  & 1.09 & 1.38 & 2.45 & 2.58 & 100.00 \\
HD 190360  & 0.82 & 1.04 & 1.84 & 1.94 & 85.67 \\
HD 192310  & 0.48 & 0.61 & 1.10 & 1.16 & 19.14 \\
HD 209100  & 0.38 & 0.48 & 0.87 & 0.92 & 100.00 \\
HD 217987  & 0.16 & 0.20 & 0.38 & 0.40 & 100.00 \\
HD 219134  & 0.41 & 0.52 & 0.95 & 1.00 & 99.26 \\
  \enddata
\end{deluxetable}

To quantify the dynamical simulation results on a system-by-system
basis, we calculate and provide the DVHZ in each case, as described in
Section~\ref{sim}. These calculations are shown in
Table~\ref{tab:results}, along with the inner and outer boundaries for
the CHZ and OHZ, described in Section~\ref{hz}. For 11 of the systems,
DVHZ is less than 50\%, and there are 4 cases where the DVHZ is 0\%,
including the previously discussed systems of HD~39091 (pi Men) and
HD~147513. In general, the architectures for the 11 cases where DVHZ
is less than 50\% are dominated by a giant planet in an eccentric
orbit that passes near or through the HZ.

\begin{figure}
  \includegraphics[width=8.5cm]{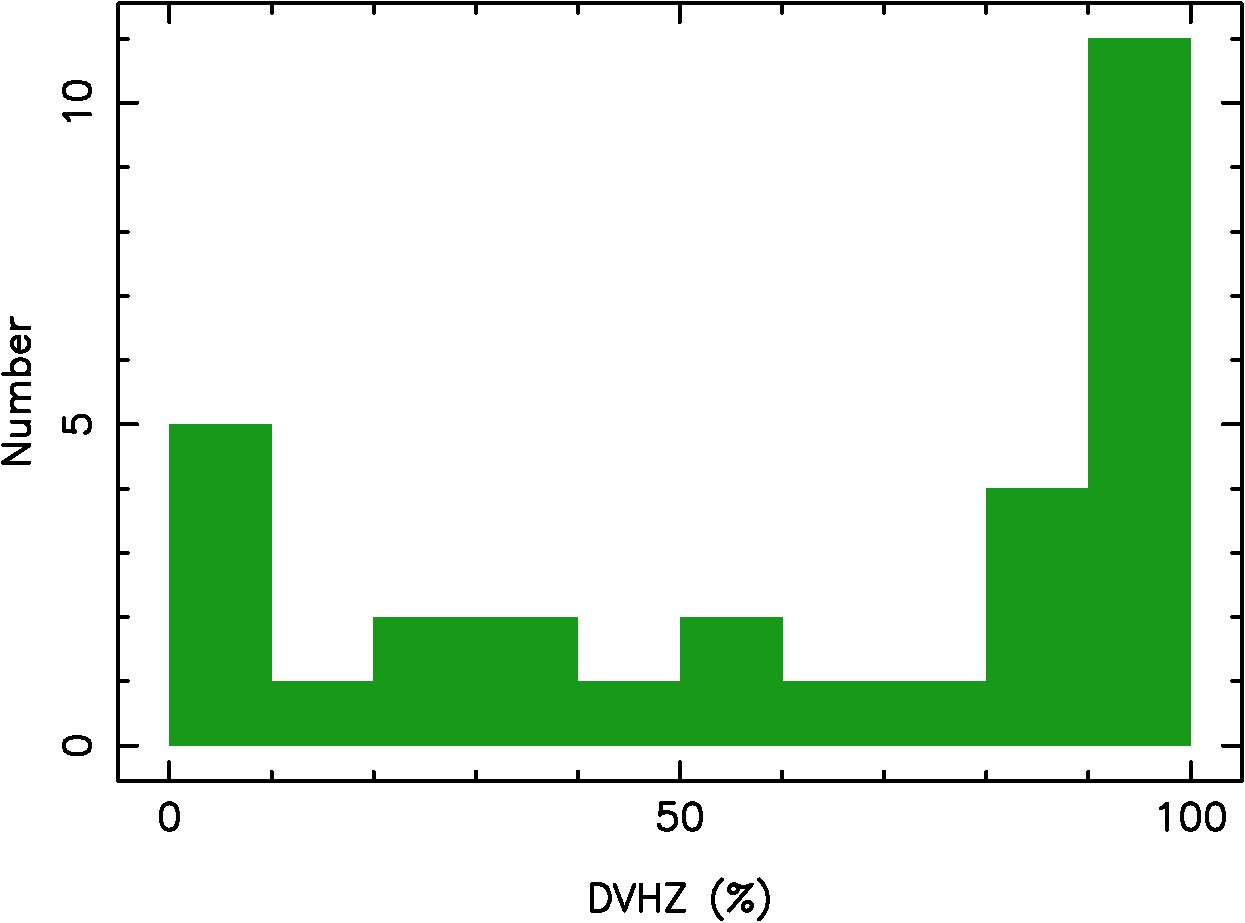}
  \caption{Histogram of the DVHZ calculations for all 30 of the known
    systems studied in this work.}
  \label{fig:dvhz}
\end{figure}

On the other hand, there are 9 cases for which the DVHZ is
100\%. Cross-referencing these systems with Table~\ref{tab:params}
reveals that their architectures are dominated by low-mass,
short-period planets with limited gravitational influence within the
HZ. Shown in Figure~\ref{fig:dvhz} is a histogram of the DVHZ values
that summarizes the findings of our dynamical simulations. This shows
that half of the studied systems have DVHZ values that are greater
than 80\%, ensuring that there is substantial dynamically viable space
within the HZ of those systems based on the currently known
architectures, and therefore suitable for potential direct imaging
follow-up observations.

\section{Discussion}
\label{discussion}

Given the number of known exoplanetary systems, including those
resulting from numerous decades of RV observations around bright
stars, it is little surprise that the HWO target list contains known
exoplanet systems. The description of these 30 systems in
Section~\ref{sample} shows that the architectures of these systems is
diverse, including low-mass multi-planet systems and high-mass
single-planets in eccentric orbits. The latter category includes giant
planets beyond the snow line, the frequency of which appears to be
relatively rare
\citep{albrow2001c,wittenmyer2011a,wittenmyer2016c,fulton2021,rosenthal2021,bonomo2023},
and can greatly influence the volatile delivery within the system
\citep{raymond2014d,raymond2017b,venturini2020b,kane2024a}. The high
eccentricity cases may be the result of planet-planet scattering
events \citep{carrera2019b} and can dynamically eliminate the
possibility of further planets being present in the system
\citep{brewer2020}. The growing understanding of such dynamically
volatile systems, including their formation and prevalence, will
greatly inform our ability to infer the possible presence of HZ
exoplanets \citep{barnes2004b}.

As described in Section~\ref{sim}, the integration time for each of
our simulations was $10^7$~years. The integration time was chosen to
minimize the effects of orbital element error propagation
\citep{ford2005a,kane2009c}, and to create a feasible pathway to
explore the full parameter space for all 30 targets, each of which
required thousands of simulations. However, chaotic systems are often
the result of divergent orbital eccentricities that can require
significant time to develop. It is thus possible that, even for
locations identified as stable in our simulations, instability occurs
beyond the $10^7$~year integration time
\citep{gozdziewski2001a}. Although our criteria for stability is a
first-order approach that necessitates all planets survive the
duration of the dynamical simulation, this criteria has successfully
been validated against chaos indicators \citep{dvorak2010} and applied
to numerous exoplanetary systems
\citep{menou2003a,dvorak2003,kane2016d}.

As noted in Section~\ref{known}, the preponderance of planets in our
sample are not known to transit their host stars, whereby the true
mass of the planet may be realized. Thus, the planetary masses used
are generally minimum masses. A consequence of this knowledge
limitation is that the dynamical influence of the planets on each
other within a system, and their effects on the injected planet, may
be larger than that calculated and presented here. For that reason,
the results presented in Section~\ref{results} can be considered a
lower limit on the HZ orbital instability. Furthermore, the
observational bias in the RV and transit methods mentioned in
Section~\ref{known} means that the architectures of the known systems
are likely incomplete, with additional planets yet to be discovered in
these systems. {However, it is worth repeating here that we assumed
  coplanar orbits for the injected planets in our simulations
  (coplanar with planet b in the case of pi~Men). A full exploration
  of terrestrial HZ planets whose orbits are mutually inclined from
  the known planets may yet reveal stable locations that are otherwise
  unviable.}


\section{Conclusions}
\label{conclusions}

The effective transition of the exoplanet community to prioritizing
direct imaging techniques requires that the data for the expected
stellar targets be sufficiently leveraged toward a robust target
list. Of the 164 targets provided by \citet{mamajek2024} for direct
imaging with HWO, only $\sim$20\% are presently known to harbor
exoplanets. Most, if not all, of the remaining stars likely also have
planetary companions, but in most cases the current database of RV
observations may be sufficient to rule out the presence of giant
planets in or near the HZ. For those that are known exoplanet hosts,
our results show that 11 of the systems have less than 50\% of the HZ
that is dynamically viable (DVHZ~$< 50$\%), including 4 cases where
there is no dynamically possible orbit (ups~And, eps~Eri, pi~Men, and
62~G.~Sco). These systems should therefore be treated with extreme
caution when considered as possible HWO targets. However, over half of
the tested systems have DVHZ values greater than 80\%, making them
suitable HWO targets from the perspective of orbital
dynamics. Moreover, our overall results are not necessarily
representative of what may be expected for the broader list of HWO
targets where no exoplanets have yet been detected. Giant planets are
relatively rare among the exoplanet population, and the detection
methods utilized are biased toward detecting those kinds of planets
more quickly than lower-mass planetary counterparts. This means that
current exoplanet surveys may have already detected many of the nearby
systems that contain significant perturbing agents (giant planets) for
possible terrestrial HZ planets.

With respect to exoplanet architectures, the refinement of the HWO
target sample will greatly benefit from the continued search for
exoplanets. The era of extreme precision RVs \citep{fischer2016}
provides opportunities to detect previously undiscovered exoplanets
around the nearby stars. These exoplanet detections will include
low-mass terrestrial exoplanets and high-mass planets with
low-inclinations relative to the plane of the sky. Additional planets
are also likely to be found in the known systems, further constraining
the stability profiles for the HZ region, as described in this
work. Results from the Gaia mission \citep{brown2021} will aid in the
detection of long-period giant planets and allow the true mass of
known RV giant planets to be determined. All of these precursor
observational efforts will continue to form an essential component for
the success of the Astro2020 recommendations, and maximize the yield
of HWO and the characterization of HZ terrestrial planets.


\section*{Acknowledgements}

We acknowledge support from the NASA Astrophysics Decadal Survey
Precursor Science (ADSPS) program under grant
No. 80NSSC23K1476. C.K.H. acknowledges support from the National
Science Foundation (NSF) Graduate Research Fellowship Program (GRFP)
under Grant No. DGE 2146752. This research has made use of the NASA
Exoplanet Archive, which is operated by the California Institute of
Technology, under contract with the National Aeronautics and Space
Administration under the Exoplanet Exploration Program. This research
has also made use of the Habitable Zone Gallery at hzgallery.org. The
results reported herein benefited from collaborations and/or
information exchange within NASA's Nexus for Exoplanet System Science
(NExSS) research coordination network sponsored by NASA's Science
Mission Directorate.


\software{REBOUND \citep{rein2012a}}




\end{document}